\documentstyle[twocolumn,aps,epsfig]{revtex}

\begin{document}

\title{Suppressed Black Hole Production from Minimal Length}

\author{S.~Hossenfelder\thanks{sabine@physics.arizona.edu}
}

\address{Department of Physics\\ 
University of Arizona\\
1118 East 4th Street\\ 
Tucson, AZ 85721, USA}

\maketitle

\noindent
\begin{abstract}
Large extra dimensions lower the Planck scale to values soon accessible. 
Motivated by String Theory, the models of large extra dimensions predict a vast number of
new effects in the energy range of the lowered Planck scale, among them the production of TeV-mass 
black holes. But not only is the Planck scale the energy scale at which effects of modified gravity become important. String Theory as well as non-commutative quantum mechanics suggest that the Planck length acts a a minimal length in nature, providing a natural ultraviolet cutoff and a limit to the possible resolution of spacetime. The minimal length effects thus become important 
in the same energy range in which the black holes are expected to form.

In this paper we examine the influence of the minimal length on the expected production rate of
the black holes.
%\vspace{1cm}
\end{abstract}

\section{Extra Dimensions}
The study of models with  Large eXtra Dimensions ({\sc LXD}s) has recently received a great deal of attention. 
These models,
which are motivated by String Theory\cite{dienesundso}, provide us 
with an extension to the Standard Model in which
observables can be computed and predictions for tests beyond the 
Standard Model can be addressed. This in
turn might help us to extract knowledge about the underlying 
theory. The models of {\sc LXD}s successfully fill 
the gap between theoretical conclusions and experimental possibilities as the extra hidden 
dimensions may have radii large enough to make them
accessible to experiment. The need to look beyond the Standard Model infected many
experimental groups to search for such Standard Model violating processes, for a 
summary see e.g. \cite{Azuelos:2002qw}.  

Arkani-Hamed, Dimopoulos and Dvali\cite{add} proposed a solution to the
hierarchy problem by introducing $d$ 
additional compactified spacelike dimensions 
in which only the gravitons can propagate. The Standard Model particles 
are bound to our 4-dimensional
sub-manifold, often called our 3-brane. Due to its higher dimensional character, the
gravitational force at small distances then is strengthened as it goes in the
radial distance $r$ with the power ${-d-1}$ instead of the usual ${-1}$.  
This results in a lowering of 
the Planck scale to a new
fundamental scale, $M_{\rm f}$, and gives rise to the exciting possibility of 
TeV scale {\sc GUT}s \cite{Dienes:1998vh}. The radius $R$ of the extra dimensions 
lies in the range mm to $10^3$~fm for $d$ from $2$ to $7$, or the inverse radius 
$1/R$ lies in energy range eV to MeV, respectively. Throughout this paper  the new
fundamental scale is fixed to  $M_{\rm f}=1$~TeV as a representative value.

\section{Black Holes in Extra Dimensions}
Using the higher dimensional Schwarzschild-metric\cite{my}, it can be derived that the
horizon radius $R_H$ of a black hole is substantially increased in the presence of
{\sc LXD}s, reflecting the fact
that gravity at small distances becomes stronger. For a black hole of mass $M$ one finds
\begin{eqnarray} \label{ssr}
R^{d+1}_H= \frac{1}{\sqrt{\pi}}\frac{8}{d+3} \Gamma\left((d+3)/2\right) \frac{1}{M_{\rm f}^{d+1}} \frac{M}{M_{\rm f}}\quad.
\end{eqnarray}
The horizon radius for a black hole with mass $\approx$~TeV is then $\approx 10^{-3}$~fm, and thus
$R_H \ll R$ for black holes which can possibly be produced at colliders or in ultra high energetic cosmic rays ({\sc UHECR}s).

Black holes with masses in the range of the lowered Planck scale should be a subject of
quantum gravity. Since there is  no theory available yet to perform this analysis, we treat
the black holes as semi classical objects.  

Consider two partons moving in opposite
directions. If the center of mass energy of the partons, 
$\sqrt{\hat s}$, reaches the fundamental
scale, $M_{\rm f}\sim 1$~TeV,
and if the impact parameter is less than $R_H$,
a black hole with mass $M\approx \sqrt{\hat s}$
can be produced.
The total cross section for such a process
can be estimated on geometrical grounds\cite{bf}
and is of order 
\begin{eqnarray} \label{crosssection}
\sigma(M)\approx \pi R_H^2 \Theta(\sqrt{\hat s}-M_{\rm min}) \quad,
\end{eqnarray}
where $\Theta$ denotes the Heaviside function and it is assumed that black
hole formation is only possible above some minimal mass, $M_{\rm min}<\sqrt{\hat s}$, which is
of order $M_{\rm f}$. The possibility of forming these TeV-scale black holes in  the
lab, or in {\sc UHECR}s respectively, has been examined in a vast 
number of publications\cite{dim,adm,Mocioiu:2003gi}, for only to mention a few. The status has been
nicely summarized in\cite{Kanti:2004nr}.

The expression for the cross section contains only the fundamental
Planck scale as a coupling constant.
We want to mention that the given  
classical estimate of the black hole production 
cross section has been under debate\cite{Voloshin:2001fe}, 
but further investigations by \cite{Solodukhin:2002ui,Giddings2} 
justify the use of 
the classical limit. However, the topic is still under
discussion\cite{Rychkov:2004sf}.
Setting $M_{\rm f}\sim 1$TeV and $d=2$ one finds 
$\sigma \approx 1$~TeV$^{-2}\approx 400$~pb.
With this it is further found that these black holes 
will be produced at {\sc LHC} in number of $\approx 10^9$ per year\cite{dim}. 

The above cross section can be derived in String Theory approximations as well
as using the Aichelburg-Sexl metric\cite{Giddings2}. In the latter case, the Schwarzschild metric
is boosted to form two colliding shock-fronts in which trapped surfaces can be
calculated, their occurrence depending on the impact parameter. Using this ansatz it is assumed
that the shock waves can be boosted to thin fronts, thus neglecting the uncertainty
of the quantum particles. This treatment is justified as the
particles with energies $\sqrt{\hat{s}}>M_{\rm f}$ have a position uncertainty smaller than
their horizon. We will see that this feature is modified under the assumption
of a generalized uncertainty arising from the minimal length.

\section{Minimal Length}
Even if a full description of quantum gravity is not yet available, there
are some general features that seem to go hand in hand with all promising candidates
for such a theory. One of them is the need for a higher dimensional spacetime,
one other the existence of a minimal length scale. As the success of String Theory arises
from the fact that interactions are spread out on the world-sheet and do no longer take
place at one singular point, the finite extension of the string has to become important at
small distances or high energies, respectively. Now, that we are discussing the possibility of
a lowered fundamental scale, we want to examine the modifications arising from this as they
might get observable soon. If we do so, we should clearly take into account the minimal length
effects.

In perturbative String Theory\cite{Gross:1987ar,Amati:1988tn},
the feature of a fundamental minimal length scale arises from the fact that strings can not probe 
distances smaller than the string scale. If the energy of a string reaches this scale $M_s=\sqrt{\alpha'}$, 
excitations of the 
string can occur and increase its extension\cite{Witten:fz}. In particular,
an examination of the spacetime picture of high-energy string scattering shows, that 
the extension of the string grows proportional to its energy\cite{Gross:1987ar} in every order
of perturbation theory. 
Due to this, uncertainty in position measurement can never become arbitrarily small.
For a review, see\cite{Garay:1994en,Kempf:1998gk}.

The minimal length scale does not only appear within string theoretical framework but also
arises from various approaches, such as non-commutative geometries, quantum loop gravity,
non-perturbative implications of T-Duality\cite{Smailagic:2003hm} or an very interesting
{\sl gedanken experiment} using micro black holes as the limiting Planck scale\cite{Scardigli:1999jh}. 

Naturally, the minimum length uncertainty is related to a modification 
of the standard commutation relations between position and 
momentum \cite{Kempf:1994su}. With the Planck scale as high as $10^{16}$~TeV, 
applications of this
are of high interest mainly for quantum fluctuations in the early universe and for 
inflation processes and have been examined closely\cite{gup}. 

In\cite{Hossenfelder:2003jz,Hossenfelder:tbp} we used a model for the effects of the minimal length 
in which the
relation between the wave vector $k$ and the momentum $p$ is modified. We assume, no matter how 
much we increase the momentum $p$ of a particle, we can never
decrease its wavelength below some minimal length $L_{\mathrm f}$ or, equivalently, we can never increase
its wave vector $k$ above $M_{\mathrm f}=1/L_{\rm f}$. Thus, the relation between the 
momentum $p$ and the wave vector $k$ is no longer linear $p=k$ but a 
function\footnote{Note, that this is similar to introducing an energy 
dependence of Planck's constant $\hbar$.} $k=k(p)$. 

For massless particles, $m=0$, this function $k(p)$ has to fulfill the following properties:
\begin{enumerate}
\item[a)]  For energies much smaller than the new scale we reproduce the linear relation: 
for $p \ll M_{\mathrm f}$ we have $p \approx k$. \label{limitsmallp}
\item[b)] It is an an uneven function (because of parity) and $k \parallel p$.
\item[c)]  The function asymptotically approaches the upper bound $M_{\mathrm f}$. \label{upperbound}
\end{enumerate} 
In general, the above properties have to be fulfilled in the limit $m \to 0$. A particle with a
rest mass close to the new scale would experience an additional uncertainty even at rest. However,
for all particles of the Standard Model it is $m^2/M_{\rm f}^2 \ll 1$ and these effects can be neglected.

The quantization in this scenario is straightforward and follows the usual procedure. 
The commutators between the corresponding operators $\hat{k}$ and $\hat{x}$ 
remain in the standard form. 
Using the well known commutation relations 
\begin{eqnarray} \label{CommXK}
[\hat x_i,\hat k_j]={\mathrm i } \delta_{ij}\quad
\end{eqnarray}
and inserting the functional relation between the
wave vector and the momentum then yields the modified commutator for the momentum 
\begin{eqnarray} \label{CommXP}
[\,\hat{x}_i,\hat{p}_j]&=& + {\rm i} \frac{\partial p_i}{\partial k_j} \quad.
\end{eqnarray} 
This results in the generalized uncertainty principle ({\sc GUP})
\begin{eqnarray} \label{gu}
\Delta p_i \Delta x_j \geq \frac{1}{2}  \Bigg| \left\langle \frac{\partial p_i}{\partial k_j} 
\right\rangle \Bigg| \quad, 
\end{eqnarray}
which reflects the fact that by construction it is not possible anymore to resolve spacetime distances
arbitrarily good. Since $k(p)$ gets asymptotically constant, its derivation $\partial k/ \partial p$
drops to zero and the uncertainty in Eq.(\ref{gu}) increases for high energies. 
The behavior of our particles thus agrees with those of the strings found by Gross and Mende as 
mentioned above.

The arising modifications derived in\cite{Hossenfelder:2003jz,Hossenfelder:tbp} can be summarized 
in the effective replacement of the
usual measure in momentum space by a modified measure which is suppressed at high
momentum
\begin{eqnarray} \label{rescalevolume4}
\frac{{\mathrm d}^{3} k}{(2 \pi)^{3}} \rightarrow \frac{{\mathrm d}^{3} p}{(2 \pi)^{3}}  
\Bigg| \frac{\partial k_{\mu}}{\partial p_{\nu}} 
\Bigg|  \quad,
\end{eqnarray}
where the absolute value of the partial derivative denotes the Jacobian determinant.
 
In the following, 
we will use the specific relation\cite{Hossenfelder:tbp} for $p(k)$ by choosing 
\begin{eqnarray}
k_{\mu}(p) &=& \hat{e}_{\mu} \int_0^{p} {\mathrm d}p'\; e^{{\displaystyle{-\epsilon ( p'^2 + m^2)}}} \label{model} \quad,
\end{eqnarray}
where $\hat{e}_{\mu}$ is the unit vector in $\mu$ direction, $p^2=\vec{p}\cdot\vec{p}$, 
%$m$ is the rest mass of the particle 
and $\epsilon=L_{\mathrm f}^2 \pi / 4 $. The factor $\pi/4$ is included to assure that for high energies the 
limiting value is indeed $1/L_{\mathrm f}$ . 
Is is easily verified that this expression fulfills the requirements (a) - (c).

The Jacobian determinant of the function $k(p)$ is best computed by adopting spherical coordinates and can be
approximated for $p \sim M_{\mathrm f}$ by
\begin{eqnarray}
\Bigg| \frac{\partial k_{\mu}}{\partial p_{\nu}} 
\Bigg| 
&\approx& e^{{\displaystyle{-\epsilon ( p'^2 + m^2)}}} \quad.
\end{eqnarray}
With this parametrization of the minimal length effects the modifications read
\begin{eqnarray} \label{gup1}
\Delta p_i \Delta x_i &\geq& \frac{1}{2}  e^{{\displaystyle{\epsilon ( p'^2 + m^2)}}} \\
\frac{{\mathrm d}^{3} k}{(2 \pi)^{3}} &\rightarrow& \frac{{\mathrm d}^{3} p}{(2 \pi)^{3}}  
e^{{\displaystyle{-\epsilon ( p'^2 + m^2)}}}  \quad. \label{gup2}
\end{eqnarray}

\section{Black Holes and the Minimal Length}
The properties of Planck size black holes raise a bunch of fundamental questions as they
exist in a regime where quantum physics and gravity are   of equal importance.
Even an examination within a not fully consistent treatment can reveal some of the exciting
and new issues on the interplay between quantum physics and gravity. One of the features
arising is the evaporation of black holes, which has first been derived in a semi classical treatment
by Hawking in 1975\cite{Hawk1} and since that time has been reproduced within various approaches.

In particular, the analysis of the last section raises the question for the final state of the black hole. This topic has been discussed in the literature extensively and is strongly connected to the 
information loss puzzle. The black hole emits thermal radiation whose sole property is its
temperature whatever the initial state of the collapsing matter has been. So if the black hole
first captures all information behind its horizon and then completely vanishes into thermally
distributed particles the basic principle of unitarity can be violated. This happens when 
the initial state was a pure quantum state and then evolves into a mixed one\cite{Hawk82}.

When we try to escape the information loss problem we have two possibilities left: the information
is released back by some unknown mechanism or a stable black hole remnant is left which keeps the
information. Besides the fact that it is unclear in which way the information should escape the horizon\cite{escape1} there are several more arguments for the 
black hole relics\cite{relics1}.

The most obvious one is the uncertainty relation. The Schwarzschild radius of a black hole with Planck mass 
is of the order Planck length. Since the Planck length is the wavelength 
corresponding to a particle of
Planck mass we see that we get in trouble when the mass of the black hole drops below 
the Planck mass. 
Then we have a mass inside a volume which is smaller than the uncertainty 
principle allows\cite{39}. For this reason is was proposed by Zel'dovich that 
black holes with masses below Planck mass should be associated with stable 
elementary particles\cite{40}. The question for black holes with regard to the
minimal length was also raises by Gross and Mende\cite{Gross:1987ar}. They found by
an investigation of the spacetime picture for such string scattering that,
with an increasing number of the order in perturbation theory, the size of
the string decreases relative to the Schwarzschild-Radius of the collision region.
The production of black holes thus does not become impossible but increasingly
difficult within the minimal length approach.
 
\section{Black Holes and the Minimal Length in Extra Dimensions}
It has been examined which modifications from the {\sc GUP} arise for the Hawking-Spectrum of the
black hole and it has been shown by Cavagli\`a, Das and Maartens\cite{Cavaglia:2003qk} that the black hole is hotter and decays
faster into a smaller number of high energetic particles, finally leaving a stable relic.
These results agree with our analysis of the 
Hawking-Spectrum using a geometrical quantization approach\cite{Hossenfelder:2003dy}.

In the following we will examine the production rates for those black holes under the
assumption of an minimal length. 

For this purpose, consider again two partons with a center of mass energy $\sqrt{\hat{s}}$ approaching head 
on in a collision. Now, their modified
uncertainty principle will smear out their focussing at energies $\sqrt{\hat{s}}>M_{\mathrm f}$. This will 
lead to an effective suppression of the black hole formation since
the probability of the partons to get trapped inside the horizon is diminished.
 
Using the {\sc GUP} formalism, we can derive this modification.
The cross section Eq.(\ref{crosssection}) assumes that the black hole captures the total energy of the 
collision and thus, the mass of the created black hole is highly
peaked around $M=\sqrt{\hat{s}}$. Due to the high rest mass of the black
hole, its remaining momentum is negligible. However, the precise mass of the black hole might be smeared out by a
form factor of order one due to energy losses during the formation and
modifications of the horizon radius by a non-zero angular momentum\cite{Giddings:2001bu}.
%\pagebreak

We will neglect this form factor and further assume the distribution 
\begin{eqnarray}
{\mathrm d}\sigma = \sigma(\sqrt{\hat{s}}) \; \delta(M-\sqrt{\hat{s}}) \; {\rm d}^3 p
\end{eqnarray}
which is easily translated into the minimal length scenario by using Eq. (\ref{gup2})
\begin{eqnarray} \label{crossgup}
{\mathrm d} \tilde{\sigma} = \sigma(\sqrt{\hat{s}}) \; \delta(M-\sqrt{\hat{s}}) e^{{\displaystyle{-\epsilon \hat{s}}}} 
\; {\rm d}^3 p \quad.
\end{eqnarray}

This can also be understood by considering the above mentioned picture of the colliding partons.
Caused by the impossibility to focus the particles, we would expect the damping to be approximately 
$R_H/\Delta x$. With $1/R_H \approx \Delta p$ and Eq. (\ref{gup1})
this yields an exponential suppression factor $\exp(- \epsilon M^2)$ for the cross section. Thus,
agreeing with the result found earlier.

The only colliders which can reach energies above the TeV-scale and therefore potentially produce
the discussed black holes are hadron colliders. To obtain the cross section for proton-proton
($pp$) collisions the partonic cross section Eq.(\ref{crossgup}) must be integrated over a folding
with the parton distribution functions ({\sc PDF}s) $f_i(x,Q^2)$. Here, the index $i$ labels the constituent partons 
of the hadron and $s=\hat{s}/xy$ is the center of mass energy of the $pp$-collision.  
\begin{eqnarray}
\frac{{\rm d}\sigma}{{\rm d}M}  
&=&  \sum_{i, j} \int_{0}^{1} {\rm d} x \frac{2 \sqrt{\hat{s}}}{x s} \times \nonumber \\
&& f_i (x,\hat{s}) f_j (y ,\hat{s})
\sigma(\hat{s}) e^{{\displaystyle{-\epsilon \hat{s}}}} \quad.
\end{eqnarray}

By definition, the {\sc PDF}s parametrize the probability of finding a parton $i$ with momentum
fraction $x$ of the hadrons momentum at a given inverse length scale $Q$ associated with the
scattering process. Usually, this scale is chosen to be the momentum transfer, that is in the $s$-channel $Q^2 \sim s$. Here, investigating the production of black holes, 
the length scale of the scattering process is limited by the Schwarzschild radius and the generalized position uncertainty\footnote{It turns out numerically that the results do not depend on this distinction.}, we thus have $1/Q \sim R_H$.
 
Further modifications for the {\sc PDF}s in the {\sc GUP} scenario, in addition to the modified
scaling in $Q$, are not to be expected. To see this, one has to keep in mind the way in which
the experimental data is extracted and further used for the common {\sc PDF}s, such as 
the {\sc CTEQ4}-Tabulars\cite{Lai:1996mg}.

%..........................................................................
\begin{figure}
\vspace*{-15mm}
\centering \epsfig{figure=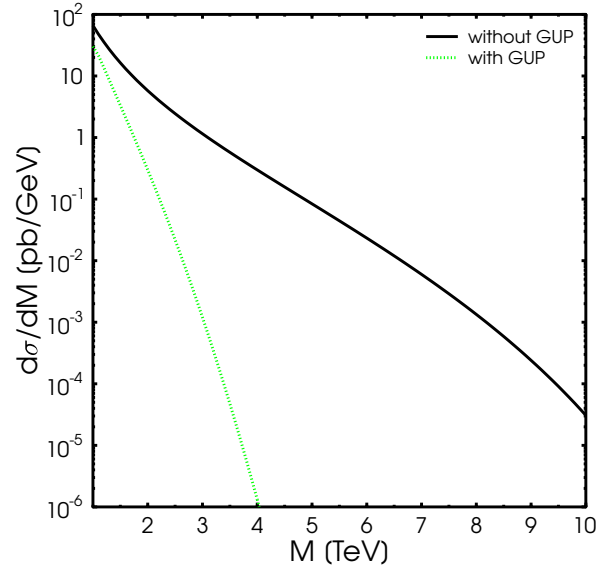,width=3.7in}
\vspace*{-5mm}
\caption{ Differential cross section for black hole production with minimal length, $1/L_{\rm f}=1$~TeV.
for {\sc LHC} energy $\sqrt{s}=14$~TeV. The differential cross section depends on $d$ only by an factor of order $1$, here $d=4$.
\label{fig1}}
\end{figure}
%..........................................................................
%..........................................................................
\begin{figure}
\vspace*{-15mm}
\centering \epsfig{figure=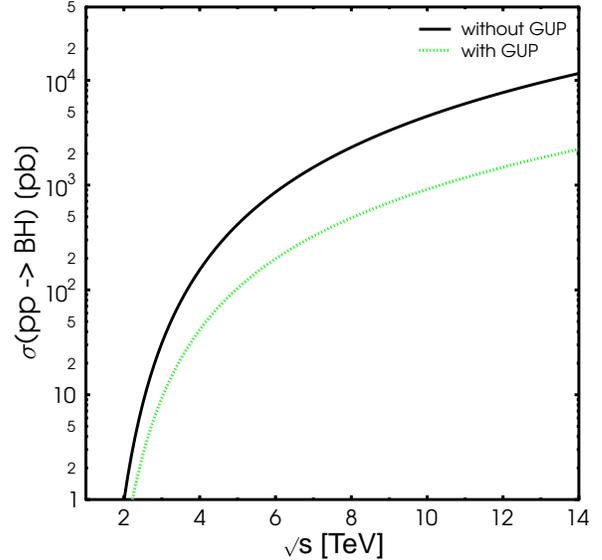,width=3.7in}  
\vspace*{-5mm}
\caption{The total cross section for black hole production with minimal length $1/L_{\rm f}=1$~TeV 
as a function of the center of mass energy $\sqrt{s}$ with $M_{\rm min}=M_{\rm f}$. The ratio of the 
total cross section with and without {\sc GUP} for the expected {\sc LHC}-energy $\sqrt{s}=14$~TeV is $\approx 0.19$.
\label{fig2}}
\end{figure}  
%.......................................................................... 

The non-perturbative physics of an hadron-hadron scattering process can 
be characterized by functions of $x$ alone at a fixed small $Q_0$ at which the minimal
length effects are negligible. This measured experimental input at small $Q_0$ is 
then extrapolated to high $Q$ using the 
{\sc DGLAP}\footnote{Dokshitzer-Gribov-Lipatov-Altarelli-Parisi} equations\cite{DGLAP}. The
scale dependence goes with $\ln ( Q^2/Q_0^2)$, signaling that $Q-$independent Bjorken scaling
is violated by {\sc QCD}-effects at high $Q^2$.

%\pagebreak

An exact analytical examination of the {\sc DGLAP} is beyond the
scope of this paper. However, since the minimal length disables a further resolution of the
hadron structure with increasing energy, the effects can effectively be captured in the above assumed $Q$-definition: 
above the new fundamental scale, the structure of the hadron is cloaked behind the 
generalized uncertainty (and it is left to the realm of philosophy to decide whether it would exist at all in that case).

The results for the above derived differential cross section and the integrated total cross-section with
use of the {\sc CTEQ4}-Tabulars\cite{Lai:1996mg} are shown in Fig.\ref{fig1} and Fig.\ref{fig2}. The calculations for the differential cross section is done for the expected {\sc LHC}-energies $\sqrt{s}=14$~TeV. 

It can be seen that the effect on the production of black holes is noticeable
but does not exceed one order of magnitude and thus stays in the range in which several other 
uncertainties might come into play (such as $M_{\rm min}$, form factors, energy losses during collapse, 
numerical factors from the analysis of trapped surfaces, $d$, angular momentum etc.). 

\section{Conclusion}

In this paper the influence of a minimal length scale on the production of black
holes in a model with large extra dimensions was examined. It was found that the
finite resolution of spacetime which is caused by the minimal length results in
an exponential suppression of the black hole cross-section. Calculation of the
total cross-section for {\sc LHC}-energies in this scenario shows a decrease of the 
expected number of black holes by a factor $\approx 5$.

\section*{Acknowledgments}

This work
was supported by a fellowship within the Postdoc-Programme of the German Academic 
Exchange Service 
({\sc DAAD}) and NSF PHY/0301998.

\end{document}